\documentclass[aps, twocolumn, superscriptaddress, footinbib, showpacs, floatfix, bibnotes]{revtex4-1}

\usepackage{amssymb}   
\usepackage{amsfonts}
\usepackage{amsmath}
\usepackage{dsfont}
\usepackage{dcolumn}   
\usepackage{bbm}
\usepackage{bm}        
\usepackage[mathscr]{eucal}
\usepackage{placeins}

\usepackage{graphicx}
\usepackage[usenames,dvipsnames]{xcolor}

\usepackage[colorlinks, linkcolor=OrangeRed,citecolor=Cyan, urlcolor=RoyalBlue]{hyperref}
\usepackage[all]{hypcap} 

\newcommand{\Le}{\left}
\newcommand{\Ri}{\right}

\newcommand{\f}{\frac}

\newcommand{\mc}{\mathcal}

\newcommand{\ra}{\rangle}
\newcommand{\la}{\langle}

\newcommand{\eq}[1]{\begin{align}#1\end{align}}

\newcommand{\msr}{\mathscr}

\newcommand{\EA}{{\msr{X}_\text{EA}}}
\newcommand{\ESG}{{\msr{X}_\text{ESG}}}
\newcommand{\ESGij}{{\msr{X}_\text{ESG}^{ij}}}

\begin{document}

\title{Accessing eigenstate spin-glass order from reduced density matrices}
\author{Younes Javanmard}
\affiliation{Max-Planck-Institut f\"ur Physik komplexer Systeme, N\"othnitzer Stra{\ss}e 38,  01187-Dresden, Germany}
\author{Soumya Bera}
\affiliation{Department of Physics, Indian Institute of Technology Bombay, Mumbai 400076, India}
\author{Markus Heyl}
\affiliation{Max-Planck-Institut f\"ur Physik komplexer Systeme, N\"othnitzer Stra{\ss}e 38,  01187-Dresden, Germany}
\date{\today}
\begin{abstract}
Many-body localized phases may not only be characterized by their ergodicity breaking, but can also host ordered phases such as the many-body 
localized spin-glass~(MBL-SG). The MBL-SG is challenging to access in a dynamical measurement and therefore experimentally 
since the conventionally used Edwards-Anderson order parameter is a two-point correlation function in time.  In this work, we show that many-body 
localized spin-glass order can also be detected from two-site reduced density matrices, which we use to construct an eigenstate spin-glass order 
parameter. We find that this eigenstate spin-glass order parameter captures spin-glass phases in random Ising chains both in many-body 
eigenstates as well as in the nonequilibrium dynamics from a local in time measurement. We discuss how our results can be used to observe MBL-SG 
order within current experiments in Rydberg atoms and trapped ion systems.
\end{abstract}
\maketitle
{\em Introduction.---}
Recently, it has been proposed that phases of quantum many-body systems may not only be characterized in terms of their 
thermodynamic properties but also on the level of single eigenstates~\cite{HuseQOrder13, FPIsing14, SondhiPRB14,Bela13,Pekker2014, Bahri2015, 
NandkishoreReview15, Parameswaran17}. 
These so-called eigenstate phases are protected by nonergodicity where the long-time dynamical properties of the system cannot 
be captured by a thermodynamic ensemble. 
Consequently, systems can exhibit order in steady states resulting from real-time dynamics although the thermal states at the corresponding energy density are featureless. 
The protecting nonergodicity can be generated by strong quenched 
disorder~\cite{Gornyi05, Basko06, Prelovsek08, Pal10, Ba12, Bela13, AltmanReview14, luitz15, Be15, BarLev15,  NandkishoreReview15, Bera17, 
AletReview17} or dynamical constraints due to gauge invariance~\cite{2017SmithD,2017SmithA,2018Brenes}. 
Recently, the dynamical signatures of such eigenstate phases have been probed in experiments including the observation of 
many-body localization (MBL)~\cite{Schreiber15, Ovadia2015, Smith2016, Choi16, Bordia16,GreinerEntg18} or discrete time 
crystals~\cite{2017ZhangObs,2017ChoiObs}.
However, some of the proposed eigenstate phases, such as the MBL spin glass (MBL-SG)~\cite{HuseQOrder13,FPIsing14, Parameswaran17}, remain challenging 
to access dynamically and thus also experimentally.

In this work we show that MBL-SG order can be detected from two-site reduced density matrices, which we use to construct 
an eigenstate spin-glass (ESG) order parameter. 
We find that this ESG order parameter captures MBL-SG phases both in eigenstates as well as in the nonequilibrium  dynamics from a \emph{local in time} measurement, which makes MBL-SG order accessible within current experiments in quantum simulators. 
In previous works MBL spin glass order in an eigenstate $|\Psi\rangle$ has been detected for spin-$1/2$ systems using an 
Edwards-Anderson (EA) order parameter~\cite{YoungRMP86,HuseQOrder13}
\eq{
	\EA = \f{1}{L^2}\sum_{i,j=1}^L \la \Psi | \sigma_i^z \sigma_j^z |\Psi \ra ^2 \, , 
	\label{eq:EA}
}
where $\sigma_i^z$, $i=1,\dots,L$, denotes Pauli matrices on site $i$ with $L$ the total number of lattice sites.
This order parameter requires access to single quantum many-body eigenstates, which experimentally is not achievable and also 
limits numerical studies to exact diagonalization and therefore the reachable system sizes. 
While $\EA$ can also be rewritten in the time domain, a measurement of $\EA$ then requires access to a two-time correlation 
function at large times, which is experimentally challenging and not possible within current quantum simulator implementations. 

\noindent
{\em Eigenstate spin-glass order parameter.---}
For the definition of the eigenstate spin-glass order parameter $\ESG$, let us first fix two lattice sites $i$ and $j$. 
Moreover, let us denote the reduced density matrix of these two sites by $\varrho_{ij}$, which can be obtained from the full 
density matrix $\varrho$ by tracing out the complement of the two sites $i$ and $j$.
The main idea behind $\ESG$ is to not calculate the square of the spin-spin correlator in Eq.~(\ref{eq:EA}) for the full quantum many-body eigenstate, but rather on the local equivalent which are the eigenstates of the reduced density matrix $\varrho_{ij}$.
Accordingly, we diagonalize the $4\times4$ matrix $\varrho_{ij}=\sum_n p_n^{if} |\psi_n^{ij}\rangle\langle \psi_n^{ij} |$ to find its 
eigenvalues~($p_n^{ij}$), eigenvectors~($|\psi_n^{ij}\rangle$) and calculate the following quantity:
\eq{
	\ESGij  = \sum_{n=1}^4 p_n^{ij} \la \psi_n^{ij} | \sigma^z_i \sigma^z_j |\psi_n^{ij} \ra ^2 \, .
	\label{eq:ESG}
}
Finally, we perform a spatial average over all pairs $(i,j)$ via:
\eq{
	\ESG  = \f{1}{L(L-1)} \sum_{i \not = j}^L \ESGij \, .
}
It is the central result of this work that for a paradigmatic MBL spin-glass model the $\ESG$ detects the eigenstate spin-glass order as we show in detail below.
Thus, MBL spin-glass order doesn't require knowledge of the full quantum many-body eigenstate, but rather only the local information contained in the reduced density matrix.
We compare $\ESG$ and $\EA$ both in the GS for large systems using DMRG and in highly-excited states using exact diagonalization.
We find numerical evidence that both of these quantities are not only quantitatively close but also can be used as order parameters for the MBL spin 
glass transition in the studied model.
Importantly, we also show that $\ESG$ can be used as a dynamical measure to detect the MBL spin-glass order.
In particular, we find that for typical initial conditions, the long-time limit of $\ESG$ is nonzero in the ordered phase and vanishes in the paramagnetic one.
However, towards the transition the dynamics becomes very slow such that accessing the structure of the transition remains challenging.
In the end we will discuss how to observe our findings in current experiments.

{\em Model and method.---}
We study the ESG order parameter $\ESG$ for the following quantum Ising chain with open boundary conditions, 
\eq{
  \msr{\hat{H}}  {=} -\frac{1}{2}\Le[ \sum_{i=1}^{L-1} J_i^z \sigma_i^z \sigma_{i+1}^z + \sum_{i=1}^{L-1} J_i^x \sigma_i^x 
  \sigma_{i+1}^x  + \sum_{i=1}^{L} h_i^x \sigma_i^x \Ri], 
\label{eq:Ham}
}
where $\sigma_{i}^{x,z}$, $i=1,\dots, L$ are the Pauli matrices and $L$ denotes the total number of lattice sites.
All the parameters appearing in this model are random and taken from 
uniform distributions. We choose $J_i^z \in [-J, J]$ 
and $h_i^x \in [-h,h]$ from uniform box distribution. 
For vanishing $J_i^x$ the model reduces to the transverse field Ising chain, which is integrable and exactly solvable by a 
mapping to a quadratic fermionic theory using a Jordan-Wigner transformation~\cite{Pfeuty70}. 
To make the model generic and non-integrable we  add a weak random $J_i^x \in [-\f{h}{4}, \f{h}{4}]$ term, which becomes equivalent to a 
two-particle interaction in the fermionic language and renders the model non-integrable.
The transverse-field Ising chain with $J^x_i=0$ exhibits a $T=0$ quantum phase transition from a paramagnetic $(J<h)$ state to 
a doubly degenerate spin-glass ground state $(J>h)$~\cite{sachdev_2011}. 
In order to explore the ground state physics for the interacting model at $J_i^x \not= 0$, we use density-matrix renormalization group (DMRG) techniques 
within a matrix product state formulation~\cite{SCHOLLWOCK201196} and used second order Suzuki-Trotter decomposition to 
exponentiate the unitary operator~\cite{Suzuki_1976}. This allows us to probe the phase transition for large system sizes reducing finite-size 
effects. In order to access the high energy 
eigenstates we use standard exact diagonalization and typically calculate $16$ eigenstates from the middle of the spectrum and 
perform an average of $\ESG$ over this set of states.  
At excited energies around $\sim 1000$ disorder configurations are used to perform statistical averaging of $\EA$ and $\ESG$, while in ground state 
$\sim 100$ disorder realizations are considered for averaging. Finally, for simulating the dynamics for 
large system sizes the time-evolving block decimation technique is used. 
To minimize finite-size effects, we calculate $\EA$ and $\ESG$ by averaging not over all pairs $(i,j)$ of lattice sites but  
rather restrict to those pairs with $|i-j|> 4$. Also to minimize the edge effects for such small system sizes we excluded 
the edge site contribution in $\EA$ and $\ESG$.

\begin{figure}[!htb]
  \centering
  \includegraphics[width=1\columnwidth]{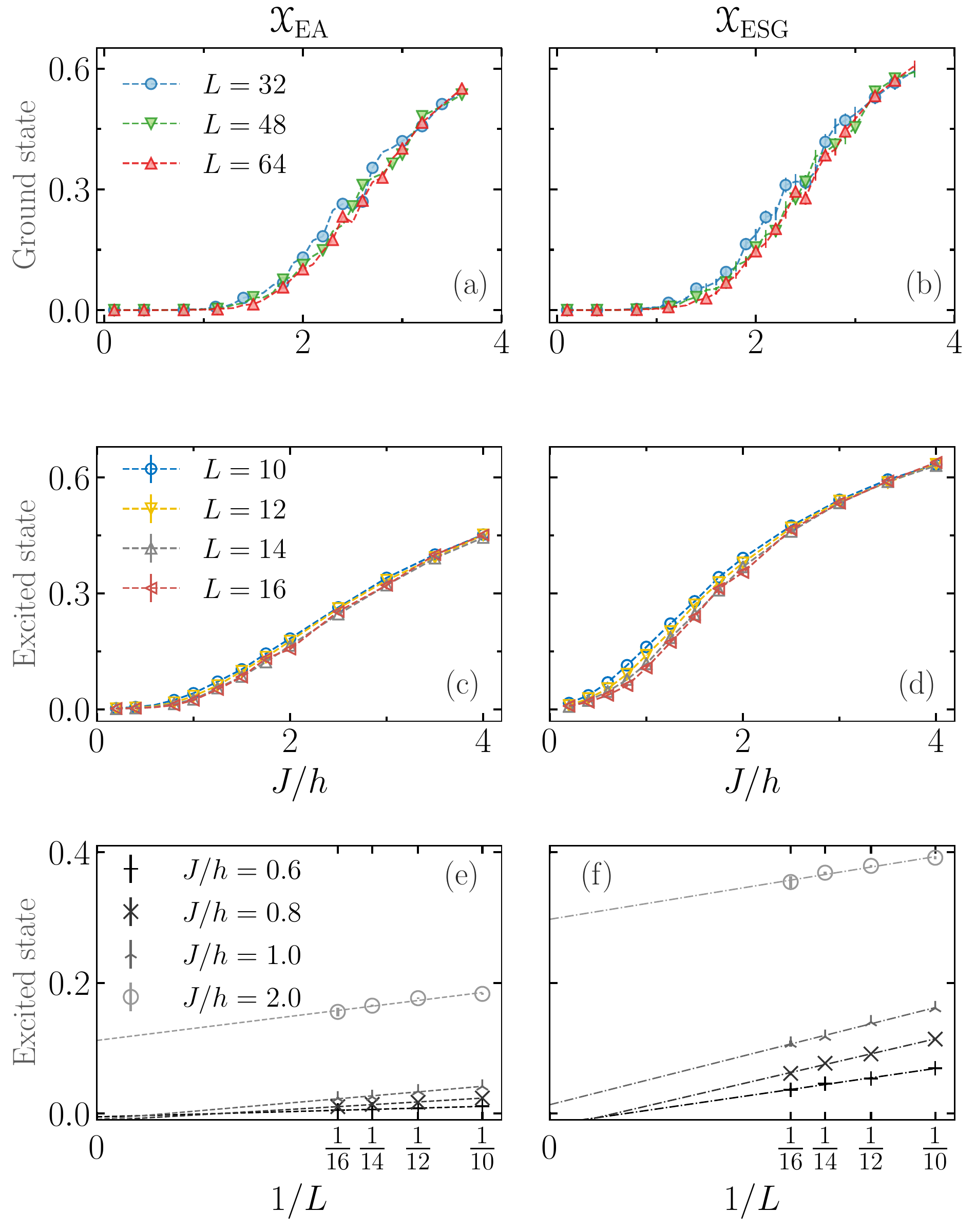}
      \caption{{\bf Eigenstate spin-glass order.} Upper panel: (a)~Shows the Edwards-Anderson order parameter, $\EA$,  in the ground state 
of the Hamiltonian Eq.~\eqref{eq:Ham} obtained using DMRG for system sizes $L={32, 40, 48}$ as a function of coupling strength $J/h$. While for weak 
couplings $\EA $ approaches zero, for large $J/h \gtrsim 1.5$ $\EA$ becomes nonvanishing in the expected spin-glass ordered phase. (b)~The eigenstate 
spin-glass order parameter $\ESG$ in the ground state for the same system sizes. (c)~$\EA$ in excited states from the center of the spectrum for 
system sizes $L={10,12,14,16}$ obtained from exact diagonalization, and compared to $\ESG$ (d). In (e-f) we show the finite-size dependence of $\ESG$ 
for excited 
states. For couplings $J/h \lesssim 1$, $\ESG$ tends towards a vanishing value for increasing system sizes, whereas for $J/h \gtrsim 1$ 
it is large and nonzero. The included lines are a guide to the eye and don't represent a quantitative extrapolation. 
  \label{fig:StaticAP}}
\end{figure}

{\em Eigenstate results.---}
First we study the $\ESG$ in eigenstates and compare it to the Edwards-Anderson order parameter $\EA$ . 
In Fig.~\ref{fig:StaticAP}(a-d) we plot both $\EA$ and $\ESG$ for the ground state calculated using the density-matrix renormalization group for system sizes $L={32, 48, 64}$ and exact diagonalization for $L={8,12,16}$ as a function of the spin-spin coupling strength 
strength $J/h$. The top panel shows the ground state results. As expected, for weak couplings~($J/h<1.5$) the 
system is in a paramagnetic phase and thus $\EA$ vanishes. The $\ESG$ is showing an analogous behavior, as can be seen in Fig.~\ref{fig:StaticAP}(b). In the spin-glass phase $\EA$ is finite and almost independent of systems size, as we find also for $\ESG$, see Fig.~\ref{fig:StaticAP}(b). Overall, these results suggest that the $\ESG$ parameter can be taken as an order parameter for the spin-glass quantum phase transition in the considered model.

In Fig.~\ref{fig:StaticAP}(c-d) we study spin-glass order in excited states of the same model Eq.~\eqref{eq:Ham} where we observe an overall similar 
behavior. 
For sufficiently weak couplings both $\EA$ and $\ESG$ take small values indicating that the system does not exhibit MBL-SG order. This is different 
for large couplings where both $\EA$ and $\ESG$ saturate to a large nonzero value almost independent of system size.
More quantitatively, we analyze the finite size dependence for strong and weak couplings by plotting both the eigenstate order parameters as a function of $1/L$ for several disorder 
values~($J/h=0.6, 0.8, 1.0, 2.5$) in Fig.~\ref{fig:StaticAP}(e-f).
While for large couplings both $\EA$ and $\ESG$ are almost independent of system size, at weak couplings a linear extrapolation in $1/L$ suggests vanishing values.
This extrapolation should be taken only as a guide to the eye as the precise functional form of the $L$-dependence is not known. 
Due to the limited system sizes and the resulting strong finite-size effects we don't attempt to extract the MBL-SG transition from the exact diagonalization data.
We find, however, that the measurement of $\ESG$ in the real-time dynamics, as discussed in the following, is much better suited for that purpose.

{\em Quench dynamics.---}
%
It is a crucial observation that $\ESG$ can be computed for any state.
In particular, we now show that this makes it possible to monitor  the buildup of MBL-SG order in the quantum real-time dynamics from a local in time measurement.
This observation is not only useful from a theoretical point of view, but also makes the observation of MBL-SG phases accessible for current experiments in quantum simulators.

In the following we study the real-time evolution of the ESG order parameter $\ESG(t)$ from initial spin configurations with random orientations along the $\sigma^x$ direction therefore respecting the $\mathbb{Z}_2$ symmetry of the Hamiltonian.
In the supplementary material we also show data for random initial spin configurations aligned along the $\sigma^z$ direction, which break the $\mathbb{Z}_2$ symmetry~\cite{supp}.
We compute the dynamics using time dependent DMRG~(tDMRG) using the second order Trotter decomposition with $dt=0.01$, where we kept only those 
states with singular values above $10^{-9}$. 
An analysis of the numerical accuracy and time steps in tDMRG calculations can be found in the supplementary material~\cite{supp}.

In Fig.~\ref{fig:dynamics_X} we show the dynamical 
evolution of $\ESG$, where we compare two representatives for the temporal behavior for weak couplings in Fig.~\ref{fig:dynamics_X}a and for strong couplings in Fig.~\ref{fig:dynamics_X}b, respectively.
At $t=0$ we have that $\ESG(t=0)=0$ since the initial condition is structureless and does not contain any spatial correlations.
In the transient stage of the dynamics we observe an increase of $\ESG(t)$ to nonzero values as a consequence of an initial buildup of spatial correlations.
On longer time scales two qualitatively different dynamical regimes emerge depending on the coupling strength, suggesting that MBL-SG order can be detected from the long-time limit of $\ESG(t)$.
While for weak couplings $\ESG(t)$ decays for increasing time, this is not the case for strong couplings, where $\ESG(t)$ saturates to a nonzero value.
%

\begin{figure}[htb]
	\centering
	\includegraphics[width=1\columnwidth]{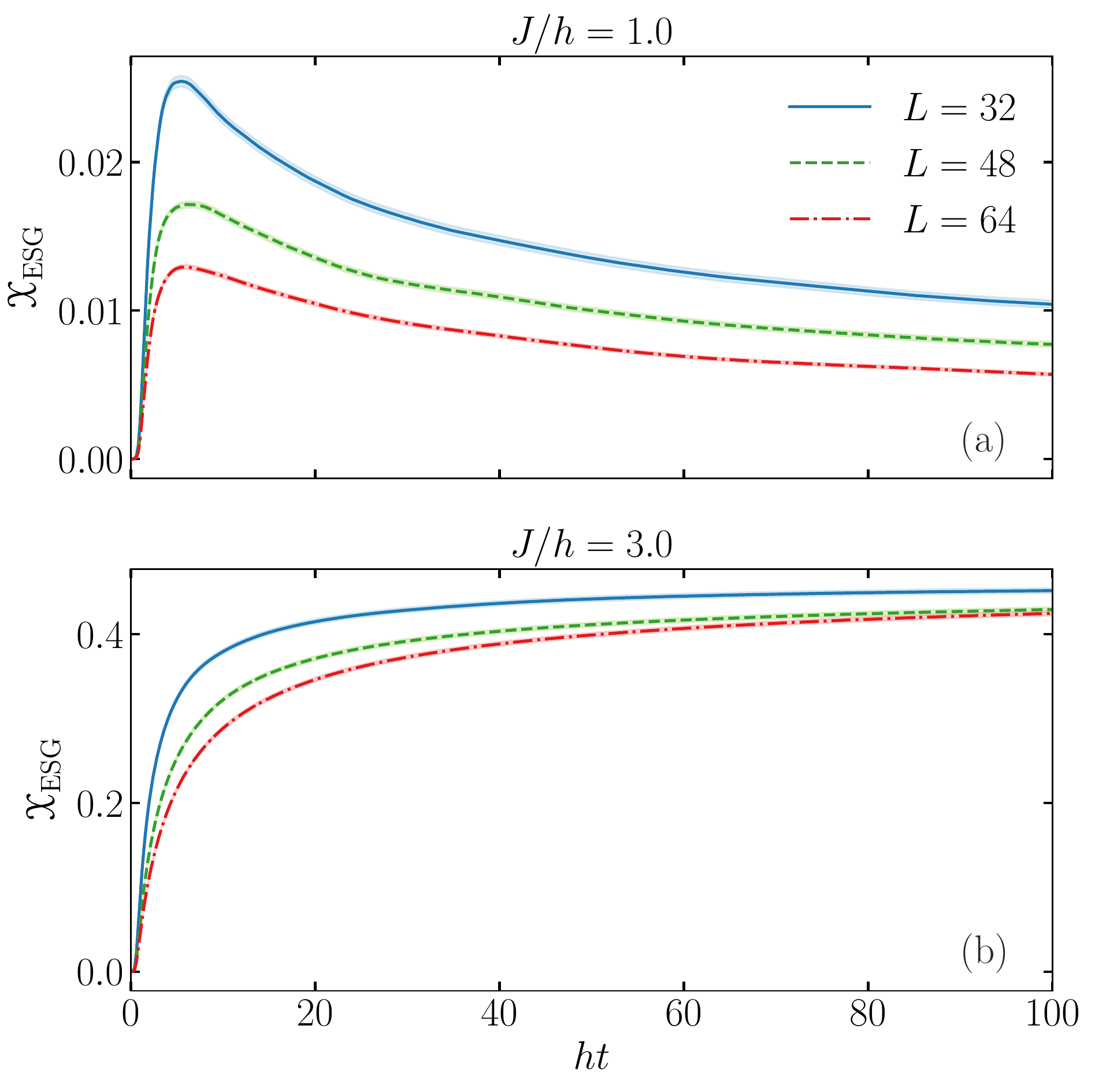}
	\caption{{\bf Quench dynamics.} Dynamical evolution of $\ESG$ 
starting from a state where all spins are initially aligned along the transverse direction. (a)~For 
		weak couplings $J/h=1.0$ $\ESG$ is small and decays on long-time scales. (b)~For larger couplings in the eigenstate spin-glass 
ordered 
phase $J/h=3.0$, $\ESG$ increases steadily with time approaching a nonzero value in the long-time limit. The shaded region indicates the statistical 
error in  the data due to a finite set of disorder averages.
		\label{fig:dynamics_X}
	}
\end{figure}

\begin{figure}[tb]
	\centering
	\includegraphics[width=1\columnwidth]{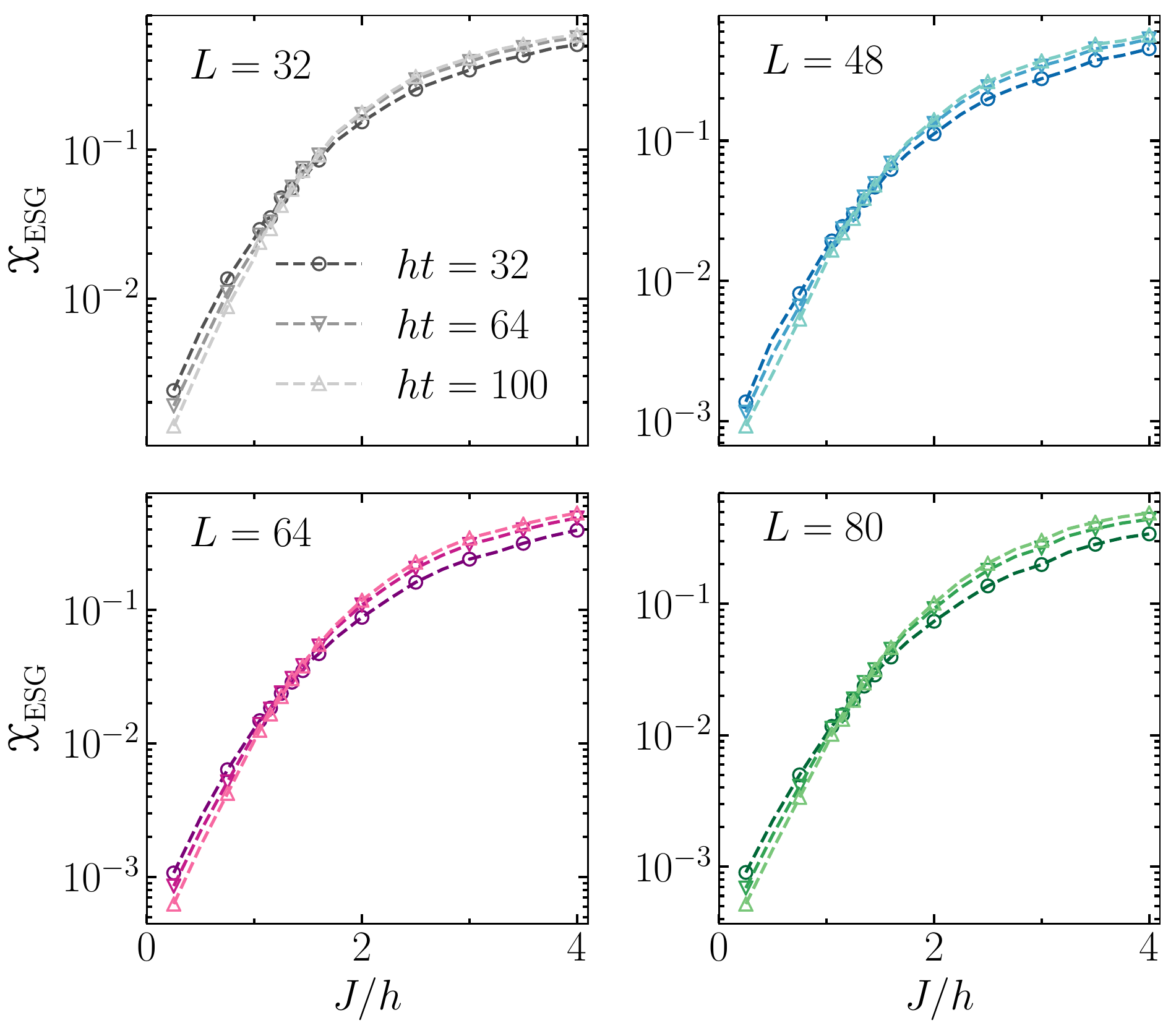}
	\caption{{\bf Eigenstate spin-glass order parameter across the transition.} Time evolution of $\ESG$ as a function of disorder strength $J/h$ for different system sizes, $L={32,48,64,80}$ in different panels. For weak couplings $\ESG$ decays with time, while 
in the opposite regime it increases. A crossing point separates these regimes of opposite dynamical behavior.
		\label{fig:dynamics_transition}
	}
\end{figure}

%
Figure~\ref{fig:dynamics_transition} demonstrates how one could possibly identify the MBL-SG transition dynamically. 
In each of the panels of this figure we plot for different system size the value of $\ESG$ as a function of the coupling strength and how this value 
changes as a function of time $t$.
For each $L$ one can identify two dynamical regimes, one where  $\ESG$ increases and one where $\ESG$ decays as a function of time within accessible 
time scales in numerical simulation.
We identify the regime where $\ESG$ goes to a nonzero value as the MBL-SG phase.
From the plots in Fig.~\ref{fig:dynamics_transition} we observe a crossing point for each system size which suggest that it might be possible to estimate the critical coupling strength for the MBL-SG transition using $\ESG$.
From our current data, however, this does not appear to be possible accurately.
For the exactly solvable case $J_i^x=0$, it is well known that the MBL spin-glass transition is located at $J/h=1$~\cite{Pfeuty70}.
Computing with tDMRG the dynamics for the same system sizes and times, we find that in this case the crossing point is located 
around $J\approx1.2$~\cite{supp}, which overestimates the region of the MBL paramagnet.
We attribute this observation to the slow expected dynamics in these models, which can show slow power-law or also logarithmic 
relaxation~\cite{MBLDynamicsRG_Altman}.
While the crossing point for the times accessible within tDMRG shows only a weak dependence on time, it is very likely that it exhibits a further slow drift on even longer times scales.

{\em Discussion.---}
In this work we have provided evidence that the detection of MBL spin-glass order does not require access on the full many-body eigenstates, as is necessary for 
the previously used Edwards-Anderson order parameter.  We rather find that MBL spin-glass order in random quantum Ising chains is contained in 
two-spin reduced density matrices. 

This observation has several implications. On the theory side, reduced density matrices can be accessed with a variety of methods, whereas full 
eigenstates require in general the use of exact diagonalization, which a priori limits the accessible system sizes. We have shown in this work, for 
example, that our proposed eigenstate spin-glass (ESG) order parameter can be computed using the density-matrix renormalization group method, which 
allows us to reach systems up to at least $80$ spins . This can be achieved, since the ESG can be obtained from a local in time measurement in 
contrast to the Edwards-Anderson order parameter, which either requires access to the full eigenstates or to a two-time correlation function.

The property, that the ESG can be obtained by a local in time measurement, not only makes MBL spin-glass order theoretically more easily accessible, but 
also makes an experimental detection more feasible. Still, the ESG requires the full reconstruction of a reduced density matrix of two spins. While 
this limits the range of applicable experimental platforms, reduced density matrices are accessible in so-called quantum simulators such as trapped 
ions~\cite{Blatt2012,Jurcevic2014}, superconducting qubits~\cite{Barends2016}, Rydberg systems, or ultra-cold atoms in optical 
lattices~\cite{Ardila2018}. The observation of MBL spin-glass order remains nevertheless challenging since the experimental 
realization of system Hamiltonians, that are capable to host MBL spin-glass ordered phases, has not yet been reported. For trapped ions systems, 
however, a way to generate an Ising Hamiltonian with random spin interactions has been proposed~\cite{Hauke2015,Grass2016}. In general, random 
interactions might also be straightforwardly realized using the digital approach to quantum simulation~\cite{Lanyon2011,Martinez2016,Barends2016} 
which is currently limited, however, by the accessible system sizes.

We have analyzed the ESG for a random quantum Ising chain, so that it is a natural question to which extent our results generalize to other models. 
While addressing this question on general grounds is beyond of the scope of this work, in the supplementary material we also show numerical results 
for a long-range Ising model with algebraically decaying spin-spin interactions, whose realization in a trapped ion system appears feasible within 
current experimental techniques. While finite-size effects in this model are stronger than for the nearest-neighbor Ising chain, we find a similar 
behavior of the Edwards-Anderson order parameter and the ESG, indicating that our results extend beyond the particular quantum Ising chain studied 
here.

{\em Acknowledgments.---}
We would like to thank J.~H.~Bardarson and Tal{\'i}a L.~M. Lezama for several discussions. 
SB acknowledges support from DST, India, through Ramanujan Fellowship Grant No. SB/S2/RJN-128/2016, and MH by the  Deutsche  
Forschungsgemeinschaft  via  the  Gottfried Wilhelm Leibniz Prize program. SB also acknowledges hospitality of the Max-Planck Institute for the 
Physics of Complex Systems, Dresden, Germany during the completion of the work.

\bibliography{MBL_BIB}

\setcounter{equation}{0}
\setcounter{figure}{0}
\setcounter{table}{0}
\makeatletter
\renewcommand{\theequation}{S\arabic{equation}}
\renewcommand{\thefigure}{S\arabic{figure}}
\renewcommand{\bibnumfmt}[1]{[S#1]}
\makeatother
\clearpage 
\widetext

\section*{Supplementary Information: Accessing eigenstate spin-glass order from reduced density matrices}

In this supplementary document we provide additional data and describe the required error analysis for the real time evolution using the time 
evolving block decimation algorithm.

\section{Real time evolution using time evolving block decimation~(TEBD) technique}
to 
We use the standard TEBD technique~\cite{SCHOLLWOCK201196} to perform unitary real time evolution starting from a given initial state. 
\begin{equation}
\label{time_evolution}
| \psi(t)\rangle = U(t) | \psi_{in}(0)\rangle = e^{-iHt} | \psi_{in}(0)\rangle, 
\end{equation}
where $H$ is the Hamiltonian of the system considered. The TEBD algorithm relies on the Suzuki-Trotter decomposition of the time-evolution operator 
$U(t)$, Eq.~\ref{time_evolution}. In this context one decomposes the time-evolution operator into $N$ small time steps, where $N$ is 
a large enough number such that the time interval $dt=\frac{t}{N}$ is small compared to the physical time scale of the system.
\begin{equation}
\label{decomposition}
	U(t) = U(dt=t/N)^N. 
\end{equation}
Due to the local nature of the Hamiltonian, i.e., only nearest-neighbor interaction, the Hamiltonian can be conveniently decomposed into sum 
of many local terms $h_i$ with support only on lattice sites $i$ and $i+1$. 
Therefore $U(dt)$ can be approximated by an $n$th-order Trotter decomposition~\cite{Suzuki_1976}. In our 
calculations we use a second order Suzuki-Trotter decomposition, where the local terms can be broken down to product of terms in even and odd sutes,
\begin{equation}
	\label{2nd order st}
	U(dt) = \prod_{\substack{i \\ even}}U_i \Le( \frac{dt}{2}\Ri)\prod_{\substack{i \\ odd}} U_i(dt) \prod_{\substack{i \\ 
even}} U_i \Le(\frac{dt}{2}\Ri) + O(dt^3). 
\end{equation}
Where the $U_n(dt)$ are the infinitesimal time-evolution operators $\exp(-ih_{i,i+1} dt)$ on the bonds $i$ which can be even or odd.

\subsection{Error sources and analysis}
\begin{figure}[thb!]
	\centering
	\includegraphics[width=0.4\columnwidth]{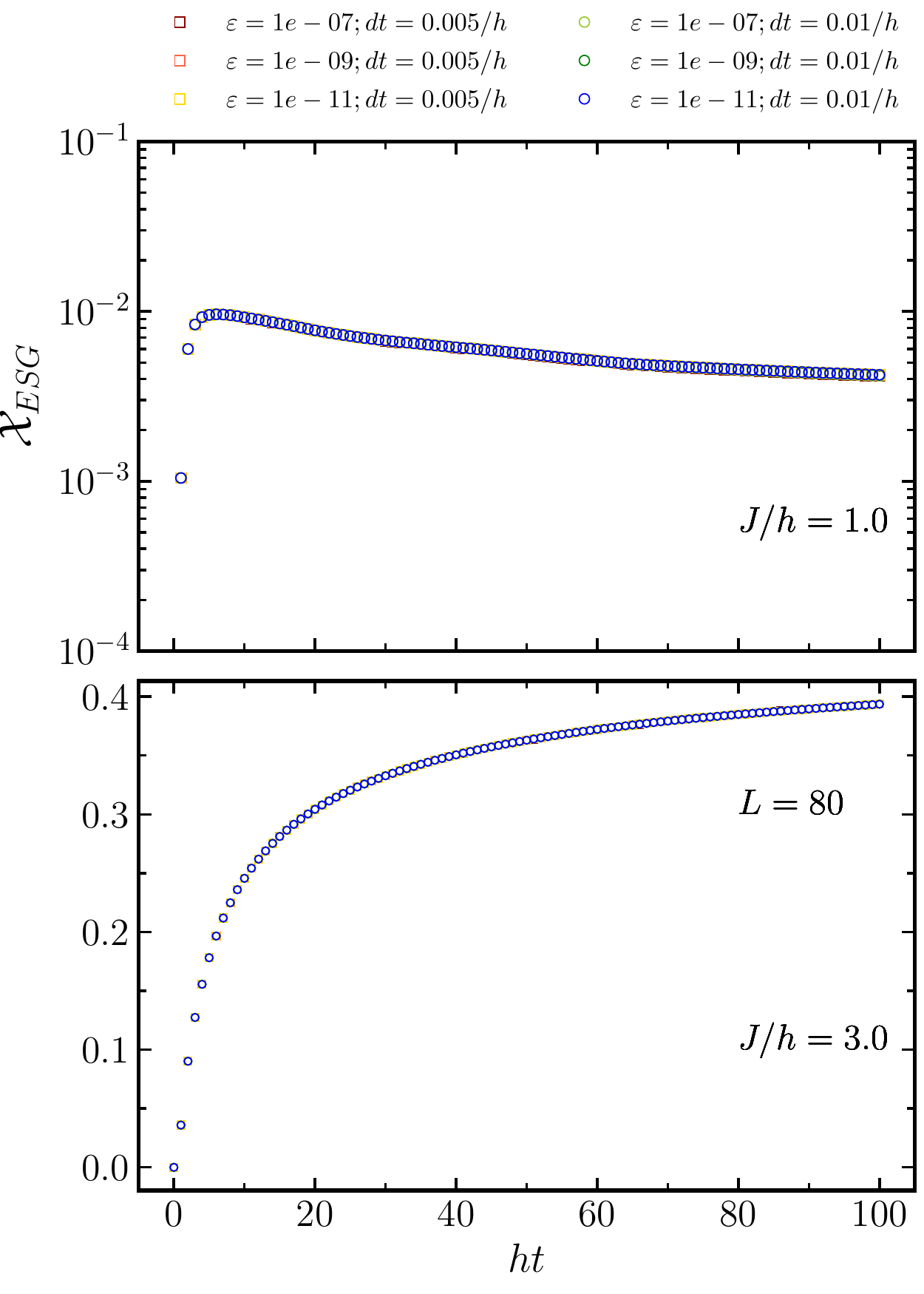}
	\includegraphics[width=0.4\columnwidth]{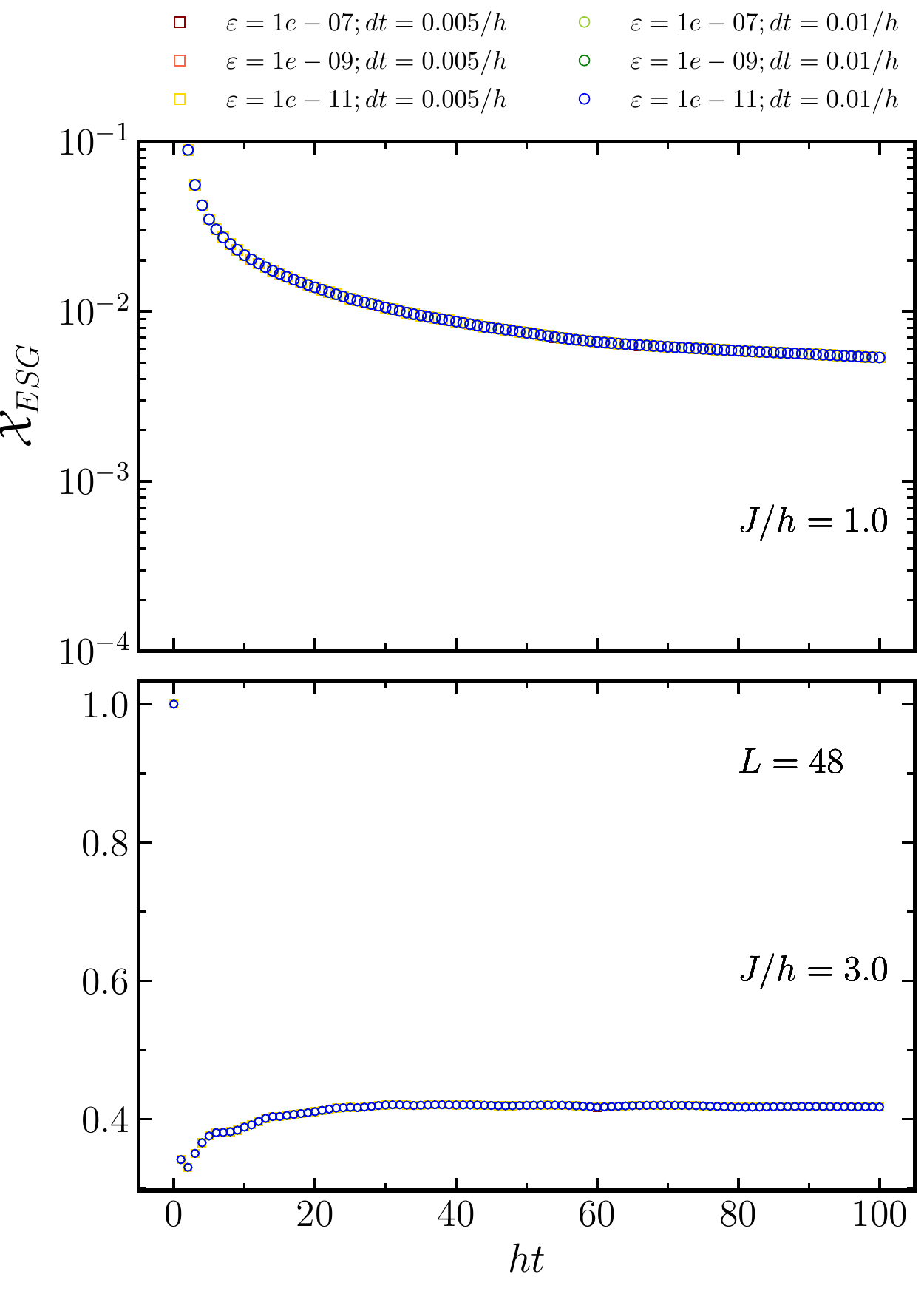}
	\caption{Error analysis for different set of parameters $(\varepsilon, dt)$ for non-MBL SG, $J/h = 1.0$, and MBL-SG, $J/h = 3.0$, phases. Left 
pannel for a chain $L = 80$ with all initial states randomly pointed in the $x$-direction. Right pannel for a chain $L = 48$ with all initial states 
randomly pointed in the $z$-direction.}
	\label{fig:error_analysis: x-dir_err}
\end{figure}
There are two main sources of error that one needs to take into account while analysing the time-dependent data. 

\paragraph{\textbf{Trotter error:}}~Due to finite degree of Trotter decomposition an error accumulates in time evolution. For an $n$th-order 
Trotter decomposition, the error in one time step $dt$ is of order $dt^{n+1}$. To reach a given time $t$, one has to perform $t/dt$ time steps, such 
that in the worst case the error grows linearly in time $t$ and the resulting error is of order $(dt)^nt$.  As well in our computation the error 
scales linearly with system size $L$, which is due to $2 \times L/2$ number of Trotter gate operation for given $dt$. Therefore the overall error is 
of the order $(dt)^nLt$. 

\paragraph{ \textbf{Truncation error:}}~This error arises due to truncation of local matrices due to ever growing Hilbert space during the TEBD run. 
The truncation error $\epsilon$  at each time step is small but it accumulates as $O(Lt/dt)$ for total time $t$. This is due to the fact that 
truncated wave function after each time step $dt$ has a norm less than $1$ and needs to be renormalized by a factor of $(1-\epsilon)^{-1} > 1$. 
Truncation errors accumulate roughly exponentially with an exponent of $\epsilon Lt/dt$ and eventually the calculations break down at very 
large time limit due to error propagation. Therefore carefully checking the error at each step and accommodating sufficient number of states to 
control it is necessary. 

\paragraph{\textbf{Optimal choice of TEBD parameters:}} In order to reach long simulation time $t$ one has to find optimal control parameters, 
which are time step $dt$, and the number of the truncated states (kept state) $\chi_{max}$. We implemented the TEBD algorithm is such a way that we 
 discarded states below certain threshold, $\varepsilon$.  

Therefore the control parameters are the time step $dt$ and the truncation error threshold is $\varepsilon$. The total error would increase at 
larger 
$dt$ due to the Trotter error, and at smaller $dt$ due to the truncation error. 
It is reasonable to choose for small times rather small values of $dt$ in order to minimize the Trotter error and for large times, to choose a 
somewhat coarser time interval, in order to push the time to as large as possible~\cite{PhysRevE.71.036102}.
We choose two small value for $dt$ 
\begin{equation}
	h dt \in [0.01, 0.005]
\end{equation}
and for $\varepsilon$ we consider different truncation thresholds 
\begin{equation}
	\varepsilon \in [1e-7, 1e-9, 1e-11]
\end{equation}

In the Following  we do the error analysis for two different initial states, first error analysis for the results that we have shown in the main text 
of the paper and second for initial state that break the $Z_2$ symmetry of the system in the $z$-direction, i.e., an initial state where all spins 
are pointed randomly in the $z$-direction. 

Figure~\ref{fig:error_analysis: x-dir_err} left panel shows the error analysis for the initial states, which is product state randomly directed in 
$x$-direction.
Note that we average over $1000$ realizations for this simulation. In each set of TEBD parameters, i.e., $\varepsilon$ and $dt$, the initial states 
remain unchanged, which give us confident that results shown here and in the main text are well converged. 
	

Figure~\ref{fig:error_analysis: x-dir_err} right panel shows the evolution of the  $\ESG$ for different TEBD parameters starting with an initial 
state, which is randomly pointed in $z$-direction. As it is seen that for all the parameters choice of the TEBD algorithm the dynamics remain 
unaffected. Here the discarded weight remains very small of the order $~1e-16$.

\subsection{Integrable model: dynamics in the vicinity of the transition}
\begin{figure}[thb!]
	\centering
	\includegraphics[width=0.7\columnwidth]{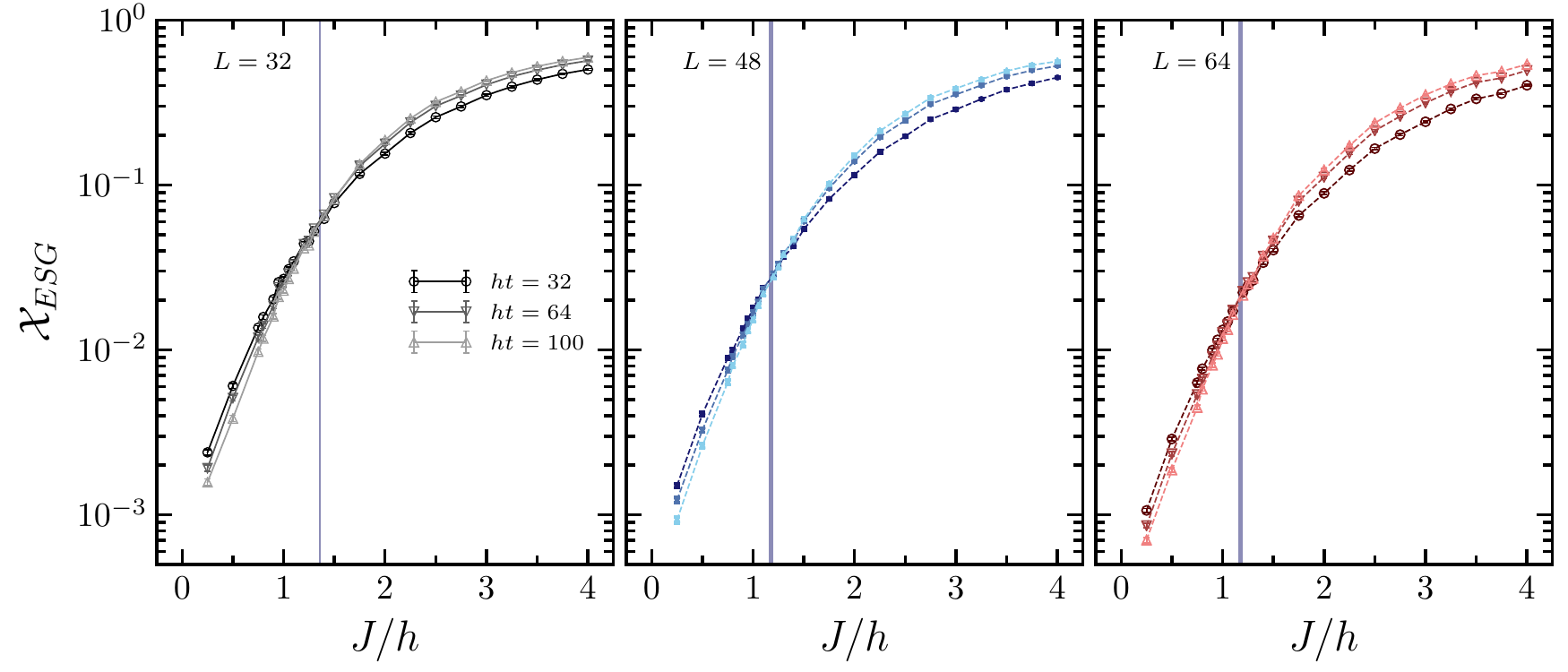}
	\caption{Eigenstate spin-glass order parameter $\ESG$ across the transition in the integrable model, where the coupling along the 
$x$-direction is taken to be zero~($J^x=0$, see main text for further details of the model). For weak couplings $\ESG$ decays with
time, while in the other regime it increases.}
	\label{fig:intg_transition}
\end{figure}

In Fig.~\ref{fig:intg_transition} we show data for $\ESG$ in the vicinity of the transition for the integrable case of the model studied in the main 
text~($J^x=0$), where the transition is known 
to be at $J/h=1$~\cite{Pfeuty70}. From this plot one can again identify two phases separated by a crossing point, with which one might identify  the 
location of the MBL-SG transition in the asymptotic long-time limit. From our finite-time data, however, we find the crossing point at $J/h\approx 
1.2$, which overestimates the transition by 20 $\%$. As in the main text, we attribute this discrepancy to the slow dynamics that can occur in these 
models, which effectively implies that one would need to reach even longer times to see a drift towards the known transition value. 

\section{Initial state with broken $Z_2$ symmetry}
\begin{figure}[h]
	\centering
	\includegraphics[width=0.55\columnwidth]{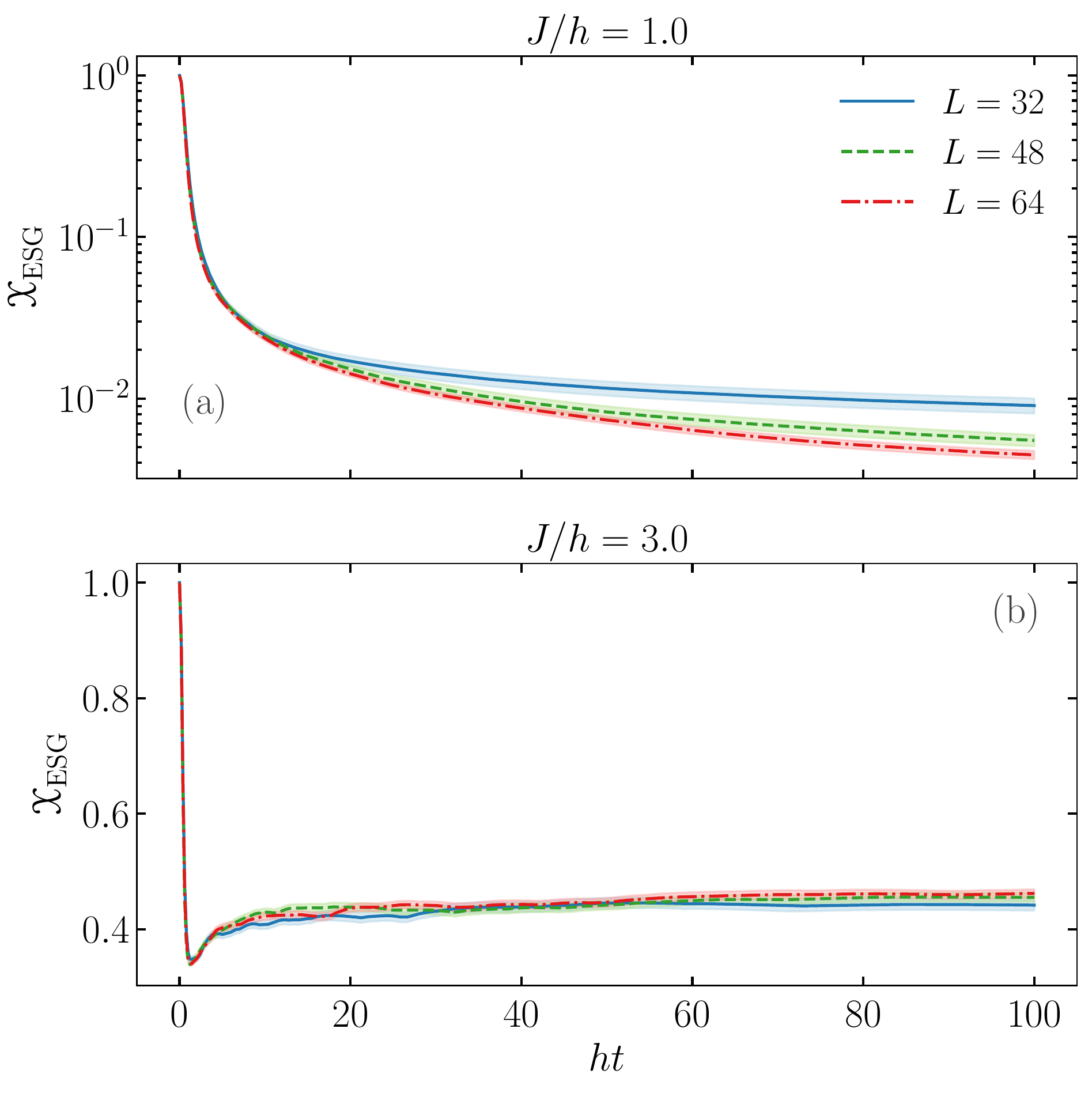}
	\caption{{\bf Quench dynamics} (a)~Shows the dynamical evolution of $\ESG$ starting from an initial 
	state where all spins initially align along the $z$- direction. Because of the initially broken $\mathbb{Z}_2$ symmetry at $t=0$, the $\ESG$ 
is non-zero as seen in both the plots. For 
	weak coupling $J/h=1.0$ the $\ESG$ decays at long times. (b)~For strong couplings $J=3.0$ the $\ESG$ becomes finite instead. The shaded region 
indicates the statistical error in 
	the data.}
	\label{fig:z-dir}
\end{figure}
In this section we show additional data for quench dynamics when the initial state breaks the Ising $\mathbb{Z}_2$ symmetry by choosing product states 
aligned along the $z$-direction. As seen in 
Fig.~\ref{fig:z-dir} we observe the same long-time dynamics as for the initial condition studied in the main text.For weak couplings the $\ESG$ values 
goes to zero with increasing system sizes at long times. For large couplings instead, $\ESG$ remains finite at 
long times for all system sizes.

\section{Long-range correlated Ising Model}
\begin{figure}[h]
	\centering
	\includegraphics[width=0.55\columnwidth]{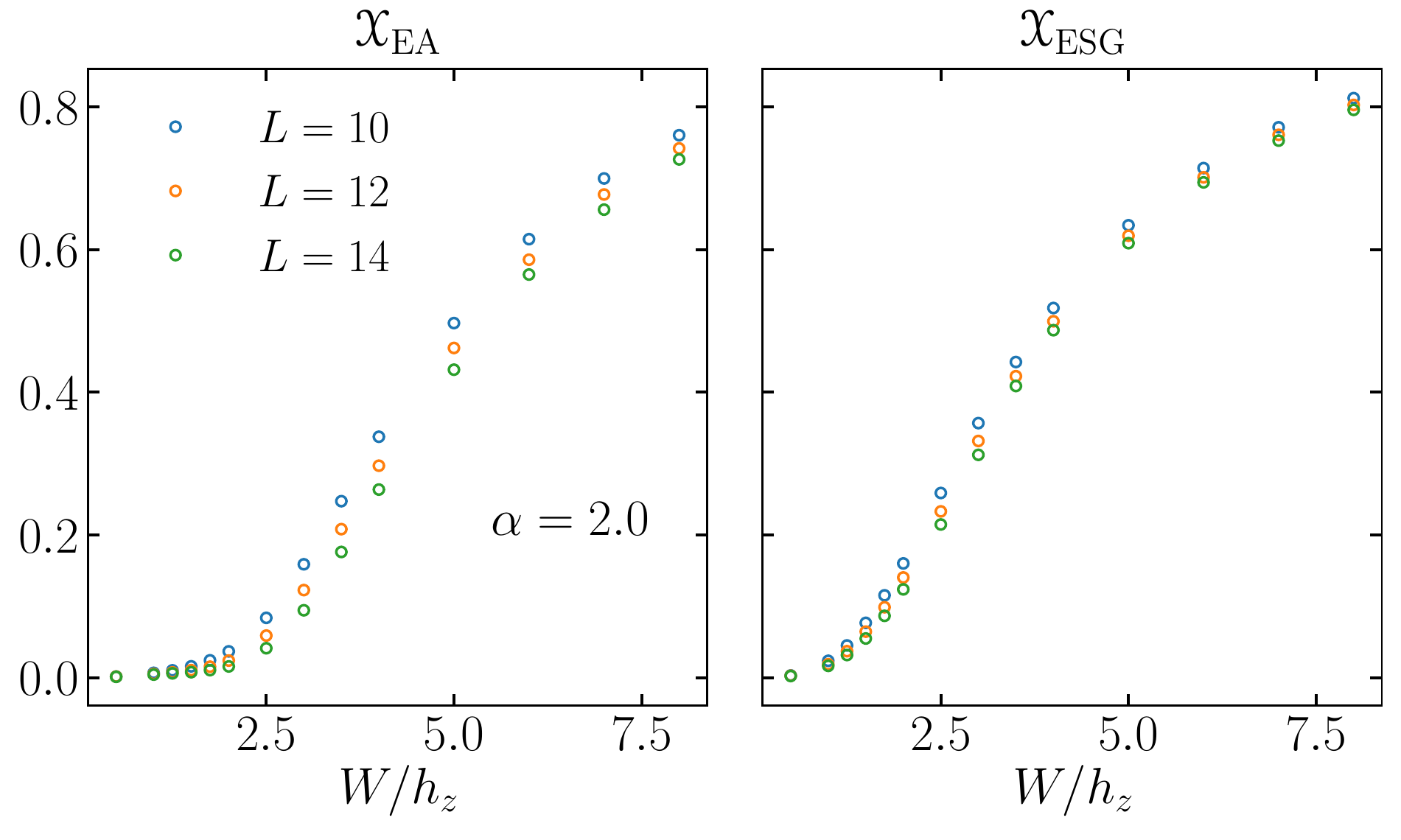}\
	\includegraphics[width=0.55\columnwidth]{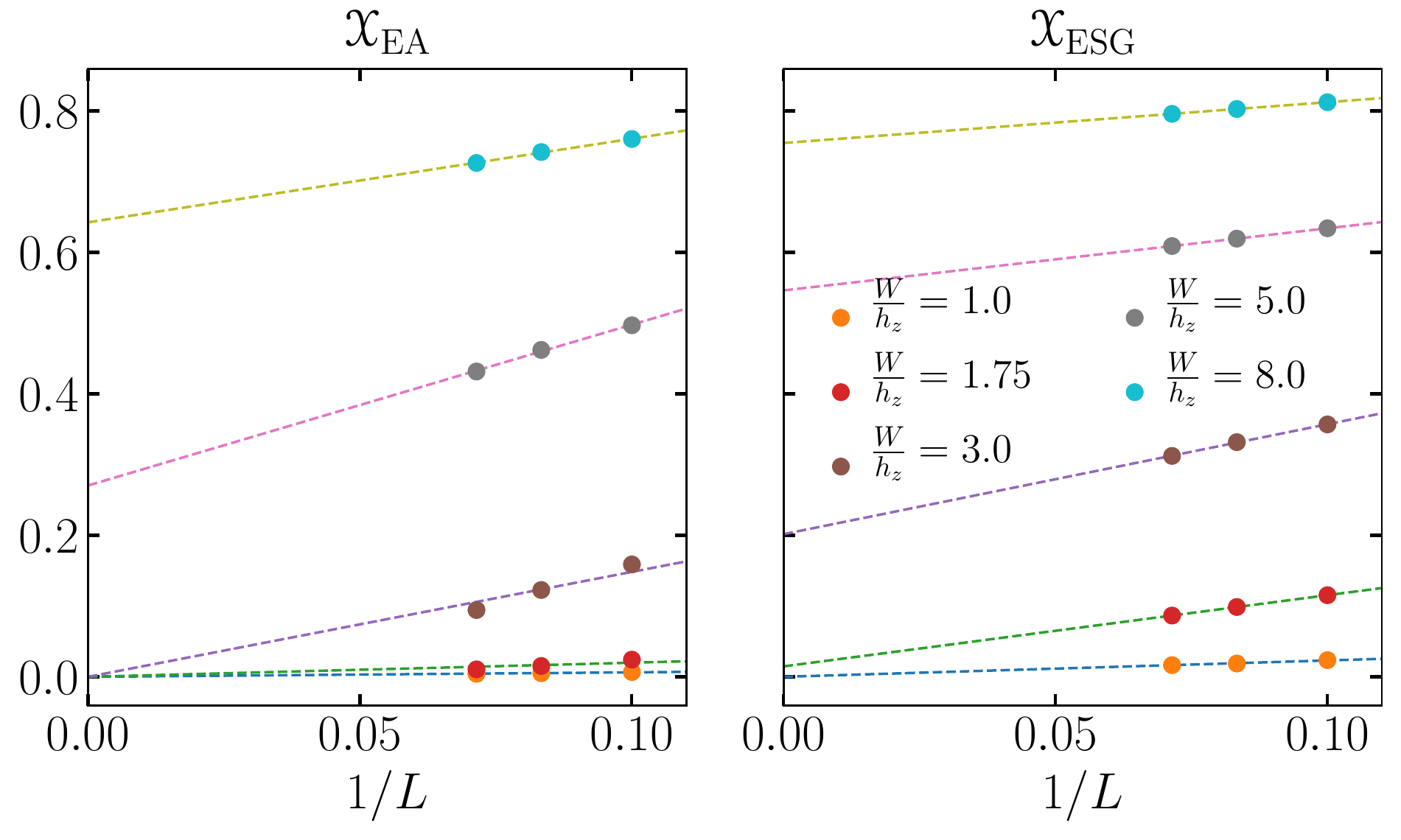}
	\caption{{\bf Eigenstate order} Upper panel: Shows both the $\EA$ and $\ESG$ for the long-range model~\eqref{eq:lrange} in for excited states 
taken from the middle of the spectrum. Above $W/h_z\gtrsim2.5$ the model appears to have a nonzero $\EA$, suggesting MBL-SG order. This 
behavior is also captured by the $\ESG$ as well. Lower panel shows the $L$ dependence of both the order parameters 
for few different values of disorder strength as mentioned in the label. Here $\alpha=2$ is chosen. As in the main text, the lines are a guide to the 
eye and are not supposed to provide a quantitative extrapolation.}
	\label{fig:longrange}
\end{figure}

In this section we consider a different model and show that also in the presence of long-range interaction the eigenstate order 
parameter shows the essential behavior of MBL-SG. The model we consider is the following:
  \eq{
    \mc{H} = \f{1}{N(\alpha)}\sum_{i<j} \frac{\Omega_i \Omega_j}{|i-j|^{\alpha}} \sigma_i^x \sigma_j^x + \sum_i h_i \sigma_i^z, 
    \label{eq:lrange}
     }
where $N(\alpha) = \sum_{i<j} \f{1}{|i-j|^\alpha}$ is the Kac normalization. $\Omega_i$ is taken from an uniform distribution $[0, W]$, while the 
field is distributed randomly between $[-h_z, h_z]$. 
Like in the main text here we also calculated both the Edwards-Anderson parameter $\EA $and the eigenstate order parameter $\ESG$. It has been 
theoretically proposed that this model can be realized in current trapped ion experiments~\cite{Hauke2015,Grass2016}.

Figure~\ref{fig:longrange} shows the parameter dependence of both $\EA$ and $\ESG$ along with the finite-size dependence. The calculation is 
performed 
for excited states taken from the middle of the spectrum for $\alpha=2.0$. As one can see the behavior of $\EA$ is well reproduced by 
the $\ESG$, however again due to strong finite size effects the critical disorder value for MBL-paramagnet to MBL-SG transition cannot be reliably 
detected. We have 
used similar averaging procedure as it is done for all the data in the main text.

\end{document}